\documentclass[twocolumn,showpacs,amsmath,amssymb]{revtex4}
\usepackage{graphicx}
\usepackage{dcolumn}
\usepackage{bm}

\makeatletter
\def\simleq{\mathrel{\mathpalette\gl@align<}}
\def\simgeq{\mathrel{\mathpalette\gl@align>}}
\def\gl@align#1#2{\lower.6ex\vbox{\baselineskip\z@skip\lineskip\z@
     \ialign{$\m@th#1\hfill##\hfil$\crcr#2\crcr\sim\crcr}}}
\makeatother

\newcommand{\fslash}[1]{\ooalign{\hfil/\hfil\crcr$#1$}}
\newcommand{\bra}{\langle}
\newcommand{\ket}{\rangle}
\newcommand{\braket}[1]{\bra #1 \ket}
\newcommand{\qq}{\braket{\bar{q}q}}
\newcommand{\sbs}{\braket{\bar{s}s}}

\newcommand{\qGq}{g\braket{\bar{q}\sigma_{\mu\nu}G_{\mu\nu} q}}
\newcommand{\sGs}{g\braket{\bar{s}\sigma_{\mu\nu}G_{\mu\nu} s}}
\newcommand{\nn}{\nonumber\\}

\newcommand{\qGqE}{g\braket{\bar{q}\sigma_{4i}G_{4i} q}}
\newcommand{\qGqB}{g\braket{\bar{q}\sigma_{jk}G_{jk} q}}

\begin{document}

\title{
Thermal effects on quark-gluon mixed condensate 
$g\langle\bar{q}\sigma_{\mu\nu}G_{\mu\nu} q\rangle$
from lattice QCD}

\author{Takumi Doi
\footnote{
Present address: 
RIKEN BNL Research Center, Brookhaven National Laboratory,
Upton, New York 11973, USA. \\
Electric address: doi@quark.phy.bnl.gov}
}
\author{Noriyoshi Ishii}
\author{Makoto Oka}
\author{Hideo Suganuma}
\affiliation{Department of Physics, Tokyo Institute of Technology, 
Ohokayama 2-12-1, Meguro, Tokyo 152-8551, Japan }


\begin{abstract}
We present the first study of the thermal effects on the quark-gluon 
mixed condensate $\qGq$, which is another chiral order parameter,
in SU(3)$_c$ lattice QCD 
with the Kogut-Susskind fermion 
at the quenched level.
Using the lattices at $\beta=6.0, 6.1$ and $6.2$ in high statistics,
we calculate $\qGq$ as well as $\qq$ in the chiral limit 
for $0 \simleq T \simleq 500{\rm MeV}$.
Except for the sharp decrease of both the condensates 
around $T_c \simeq 280 {\rm MeV}$, the thermal effects are found to be 
weak below $T_c$.
We also find that the ratio 
$m_0^2 \equiv \qGq / \qq$ 
is almost independent of the temperature
even in the very vicinity of $T_c$.
This result indicates nontrivial similarity 
in the chiral behaviors of the two different condensates.
%
%
\end{abstract}

\maketitle



\maketitle


\section{Introduction}
\label{sec:intro}

Quantum chromodynamics (QCD)
exhibits interesting nonperturbative phenomena such as
spontaneous chiral-symmetry breaking 
and color confinement, 
which are considered to be related to the 
nontrivial structure of the QCD vacuum.
In order to clarify the mechanism of these phenomena and their 
relation to the QCD vacuum, extensive studies 
have been made.
In fact, at high temperature, QCD is believed to exhibit
phase transition into the quark-gluon plasma (QGP),
where chiral symmetry is restored and the color is deconfined.
The QGP phase is considered to 
have been actually realized
in the early universe, and the on-going RHIC experiments
attempt to produce QGP
in the laboratory
through relativistic heavy-ion collisions, which motivates
further studies of finite temperature QCD.

For the theoretical study of the QCD vacuum structure 
at finite temperature, 
the condensates such as $\qq$, $\alpha_s\bra G_{\mu\nu}G_{\mu\nu} \ket$ 
and $\qGq$ play the relevant role, 
because they characterize the nontrivial QCD vacuum directly.
In fact, at finite temperature, 
changes of the structure of the QCD vacuum 
will be represented by
the thermal effects on the condensates.

In this paper,
we study the thermal effects on
the quark-gluon mixed condensate
$\qGq \equiv 
{g\braket{\bar{q}\sigma_{\mu\nu}G_{\mu\nu}^A \frac{1}{2}\lambda^A q}}$
for the following reasons.
First of all, we note that
the chirality of the quark in the operator $\qGq$
flips  as
\begin{eqnarray}
\lefteqn{
\qGq 
} \nn
&=&
  g\braket{\bar{q}_R (\sigma_{\mu\nu}G_{\mu\nu}) q_L}
+ g\braket{\bar{q}_L (\sigma_{\mu\nu}G_{\mu\nu}) q_R}.
\end{eqnarray}
Therefore, $\qGq$ 
plays the role of a chiral order parameter,
which can indicate 
chiral restoration at finite temperature, as well as 
the usual quark condensate $\qq$.
%

Secondly, 
we emphasize that $\qGq$
characterizes different aspect of the QCD vacuum
from $\qq$.
%
%
In particular, 
$\qGq$ reflects the  color-octet components
of quark-antiquark 
pairs in the QCD vacuum,
while $\qq$ reflects only the color-singlet $q$-$\bar{q}$ components.
The mixed condensate $\qGq$ thus represents
the direct correlation between color-octet $q$-$\bar{q}$ pairs
and the gluon field strength $G^A_{\mu\nu}$,
i.e., the 
color-electromagnetic field 
spontaneously generated
in the QCD vacuum~\cite{Savvidy&Nielsen}.
Therefore, 
the thermal effects on $\qGq$ 
in comparison with that of $\qq$ will 
give new and important information on 
chiral restoration of the QCD vacuum at finite temperature.

As the third point, we show how the mixed condensate
affects the properties of hadrons.
To clarify the physical meaning of the condensates, 
the QCD sum rule 
is a useful framework, in which 
the hadronic properties can be directly connected
to the various condensates with the help of the 
dispersion relation~\cite{Narison}.
One of the examples where the mixed condensate 
plays an important role is the QCD sum rule for baryons~\cite{Ioffe,Dosch}.
In the actual calculation of such sum rules, 
especially in the decuplet baryons, 
the contribution from $\qGq$ amounts to the same 
magnitude as the leading contribution from $\qq$,
and the mixed condensate has 
large effects on the $N$-$\Delta$ splitting~\cite{Dosch}.
In terms of the chiral properties of baryons, 
the parity splittings of baryons stem from the chiral-odd condensates 
such as $\qq$ and $\qGq$~\cite{Jido:parity}.
Furthermore, recent study~\cite{penta1} shows that 
the parameter $\sGs / \sbs$ 
is a key quantity for
the prediction on the parity of the recently discovered 
penta-quark baryon, $\Theta^+(1540)$~\cite{Nakano}.
%
The mixed condensate
is also important in other sum rules 
such as light-heavy meson systems~\cite{Dosch2}, 
and exotic meson systems~\cite{Latorre}.
In these respects, 
the evaluation
of the thermal
effects on $\qGq$ 
is expected to give a useful input in QCD sum rules to investigate 
hadron phenomenology at finite temperature.


For the analysis of the thermal effects on the mixed condensate $\qGq$,
we use lattice QCD Monte Carlo simulation, which 
is the direct and nonperturbative calculation from QCD.
So far, the mixed condensate at zero temperature has been 
analyzed phenomenologically in the QCD sum rules~\cite{Bel}.
In the lattice QCD, a pioneering work~\cite{K&S} was done 
long time ago, but the result was rather preliminary because
the simulation was done with insufficient statistics 
using a small and coarse lattice.
Recently, new lattice calculations have been developed
by our group~\cite{DOIS:qGq} using the Kogut-Susskind (KS) fermion,
and by another group~\cite{twc:qGq} using the Domain-Wall fermion.
At finite temperature, however, there has been no 
result on $\qGq$ except for 
our early reports~\cite{DOIS:T}.
Therefore, we present in this paper the first and intensive 
results of the thermal effects on $\qGq$ as well as $\qq$,
including the analysis near the critical temperature.

This paper is organized as follows.
In Sec.~\ref{sec:formalism},
we explain the formalism to evaluate
the condensates.
In Sec.~\ref{sec:results},
we present the lattice QCD data,
and discuss the physical implication of the results.
Sec.~\ref{sec:summary} is devoted to the summary of the paper.



\section{Formalism}
\label{sec:formalism}

In the calculation of $\qGq$ and $\qq$,
we use the SU(3)$_c$ lattice QCD with the KS-fermion 
at the quenched level.
We note that the KS-fermion preserves the explicit chiral symmetry
for the quark mass $m=0$, which is a desirable feature
to study both the condensates of chiral order parameters.
In order to evaluate the condensates, we calculate
the SU(4)$_f$ flavor-averaged condensates on the lattice as
\begin{eqnarray}
a^3 \qq
&=& - \frac{1}{4}\sum_f {\rm Tr}\left[ \braket{q^f(x) \bar{q}^f(x)} \right]
\end{eqnarray}
and
\begin{eqnarray}
\lefteqn{
a^5 \qGq
} \nn 
&=& - \frac{1}{4}\sum_{f,\ \mu,\nu}{\rm Tr}
        \left[ \braket{q^f(x) \bar{q}^f(x)} \sigma_{\mu\nu} G^{\rm lat}_{\mu\nu}(x)
\right].
\end{eqnarray}
Here, ``${\rm Tr}$'' refers to the trace over the spinor and the color
indices,  and  $\braket{q^f(y)  \bar{q}^f(x)}$ denotes  the  Euclidean
quark propagator of the $f$-th flavor.
For the gluon field strength $G^{\rm lat}_{\mu\nu}$,
we adopt the clover-type definition on the lattice
to eliminate ${\cal O}(a)$ discretization error,
\begin{eqnarray}
{G_{\mu\nu}^{A\ {\rm lat}}}
=
\frac{i}{8}
{\rm Tr}
\Bigl[ 
\lambda^A
\left(
U_{\mu\nu} +U_{\nu\,-\!\mu}+U_{-\!\mu\,-\!\nu}+U_{-\!\nu\,\mu}
\right)
\Bigr] + {\rm h.c.} , \nn
\label{eq:clover}
\end{eqnarray}
where $U_{\mu\nu}$ denotes the plaquette operator, and
$\lambda^A$ 
($A=1,2,\cdots,8$)  
is the color SU(3) Gell-Mann matrix. 
%
In the KS-fermion formalism, SU(4)$_f$ quark-spinor fields, 
$q$ and $\bar{q}$, are converted into 
single-component Grassmann 
KS-fields, $\chi$ and $\bar{\chi}$, respectively, 
with the proper insertion of gauge-link variables
which ensure the gauge covariance~\cite{DOIS:qGq}.
Note that the contraction of flavor and spinor indices
corresponds to summation over the KS-hypercube~\cite{DOIS:qGq} 
and suppresses ${\cal O}(a)$ discretization error.
The detailed 
formulations
and the diagrammatic representations
for the calculation of the condensates are given in Ref.~\cite{DOIS:qGq}.

We perform Monte Carlo simulations with the 
standard Wilson gauge action 
for $\beta \equiv 2N_c/g^2 = 6.0, 6.1$ and $6.2$.
%
%
The lattice units are obtained as
$a^{-1} =$ $1.9,$ $2.3$ and $2.7 {\rm GeV}$ for 
$\beta =$ $6.0,$ $6.1$ and $6.2$, respectively, 
where we use the lattice data~\cite{Gockeler:scale}
with the string tension 
$\sqrt{\sigma} = 427 {\rm MeV}$.

In order to perform the calculation
at various temperatures,
we use the following lattices,

\begin{tabular}{cl}
i)   & $\beta = 6.0$,\ \ $V=16^3\times N_t\ (N_t=16,12,10,8,6,4)$,\\
ii)  & $\beta = 6.1$,\ \ $V=20^3\times N_t\ (N_t=20,12,10,8,6)$, \\
iii) & $\beta = 6.2$,\ \ $V=24^3\times N_t\ (N_t=24,16,12,10,8)$,
\end{tabular}\\
which correspond to $0 \simleq T \simleq 500{\rm MeV}$ 
with the spatial volume of $L^3 \simeq (1.6-1.8 {\rm fm})^3$.
The  critical temperature  $T_c$  is obtained  from the  Polyakov-loop
susceptibility~\cite{Boyd}  in terms of  the confinement/deconfinement
phase transition.  From the  Polyakov-loop susceptibility on the above
three lattices, we  obtain $T_c / \sqrt{\sigma} \simeq  0.64$, or $T_c
\simeq$ 280MeV, which is consistent with Ref.~\cite{Boyd}.

We generate 100 gauge configurations for each lattice,
where we pick up each configuration for every 500 sweeps after 
1000 sweeps for the thermalization.
In the vicinity of the phase transition point, 
namely, $20^3\times 8$ at $\beta=6.1$ and $24^3\times 10$ at
$\beta=6.2$, the fluctuations of the condensates get larger.
We hence generate 1000 gauge configurations on these 
two lattices for the accurate estimate.
For the lattices at $T > T_c$, we only use
the gauge configurations which are continuously connected
to the trivial vacuum $U_\mu=1$.

To obtain the condensates,
we calculate the Euclidean propagator by
solving the matrix inverse equations.
Here, we use the current-quark mass of 
$ma = 0.0105,$  $0.0184$ and $0.0263$ ($\beta=6.0$), 
$ma = 0.00945,$ $0.0162$ and $0.0234$ ($\beta=6.1$) and 
$ma = 0.00770,$ $0.0134$ and $0.0192$ ($\beta=6.2$), 
which 
correspond to the physical mass of 
$m \simeq 20,35$ and $50 {\rm MeV}$, respectively.
For the KS-fields, $\chi$ and $\bar{\chi}$, the anti-periodic condition 
is imposed at the boundary in both the temporal and spatial directions.
In each configuration, we measure the condensates 
on 16 different physical space-time points at $\beta=6.0$ and 
2 points at $\beta = 6.1$ and $6.2$.
These points are taken so as to be
equally spaced in the lattice 4-dimensional volume, i.e.,
on the lattices with the volume $V=L^3\times N_t \equiv (2l)^3\times (2n_t)$
in the lattice unit,
we take 16 points at $\beta=6.0$ as $x=(x_1,x_2,x_3,x_4)$ 
with $x_i \in \{0, l\} (i=1,2,3), x_4 \in \{0,n_t\}$,
and 2 points at $\beta=6.1$ and $6.2$ as  
$x \in \{(0,0,0,0),(l,l,l,n_t)\}$.
In addition, the summation over the KS-hypercube 
is taken on each selected space-time point.
At each quark mass $m$ and temperature, 
we thus achieve high statistics as
1600 data at $\beta=6.0$ and 200 data at $\beta=6.1$ and $6.2$.
Note that for the lattices of 
$20^3\times 8$ ($\beta =6.1$)
and $24^3\times 10$ ($\beta =6.2$), 
we obtain 2000 data, which guarantees reliability of the results
even in the vicinity of the critical temperature.


At each temperature,
we calculate the condensates $\qq$ and $\qGq$
at three different quark masses, $m$.
We observe that both the condensates show 
a clear linear behavior against the quark mass
at every finite temperature,
as  was seen in Ref.~\cite{DOIS:qGq} at $T=0$.
Therefore, we fit the data with a linear function and determine 
the condensates in the chiral limit.

We comment here on the error estimate and the numerical check on the finite-volume effect in lattice QCD.
The statistical error is estimated with the jackknife method.
In the vicinity of $T_c$, i.e., on $20^3 \times 8$ at $\beta=6.1$ and $24^3 \times 10$ at $\beta=6.2$,
we conservatively suppress the autocorrelation by 
performing binning of the size of 10.
We find that the statistical error is typically 5-7\% level 
for every temperature.
For the check of the finite-volume artifact, we examine the condensates imposing 
the periodic condition on the KS-fields, $\chi$ and $\bar{\chi}$,
instead of the anti-periodic condition,
at the 3-dimensional spatial boundaries.
We perform the calculations with typical lattices of 
$24^3\times 10$ $(\beta=6.2)$ and
$16^3 \times 4$ $(\beta=6.0)$, 
which corresponds to $ T\simeq T_c$ and $T_c \ll T$, respectively,
and find that the dependence of the condensates on 
the spatial boundary condition is only 1\% level. 
For $T \ll T_c$ lattice case, 
it was shown that 
the finite-volume artifact for $16^4 (\beta=6.0)$ lattice
is only 1\% level
in Ref.\cite{DOIS:qGq}.
We therefore conclude that the physical volume
$L^3 \simgeq (1.6 {\rm fm})^3$ in our simulations 
is large enough to
avoid the finite volume artifact.


\section{The Lattice QCD Results and Discussions}
\label{sec:results}

We estimate the thermal effects
on each condensate, $\qGq$, or $\qq$, 
by taking the ratio between the values at finite 
and zero temperatures.
In this ratio, 
the renormalization constants 
are canceled
because there is no operator mixing for 
both the condensates
in the chiral limit~\cite{Narison2}.
Here, at each $\beta$, we use the value at the 
lowest temperature 
as a substitute for the value 
at zero temperature.
This approximation can be justified because 
the finite-volume artifact for the lowest temperature 
is almost negligible, as was shown for $16^4 (\beta=6.0)$ 
lattice.

\begin{figure}[htb]
\centering
\includegraphics[scale=0.24]
{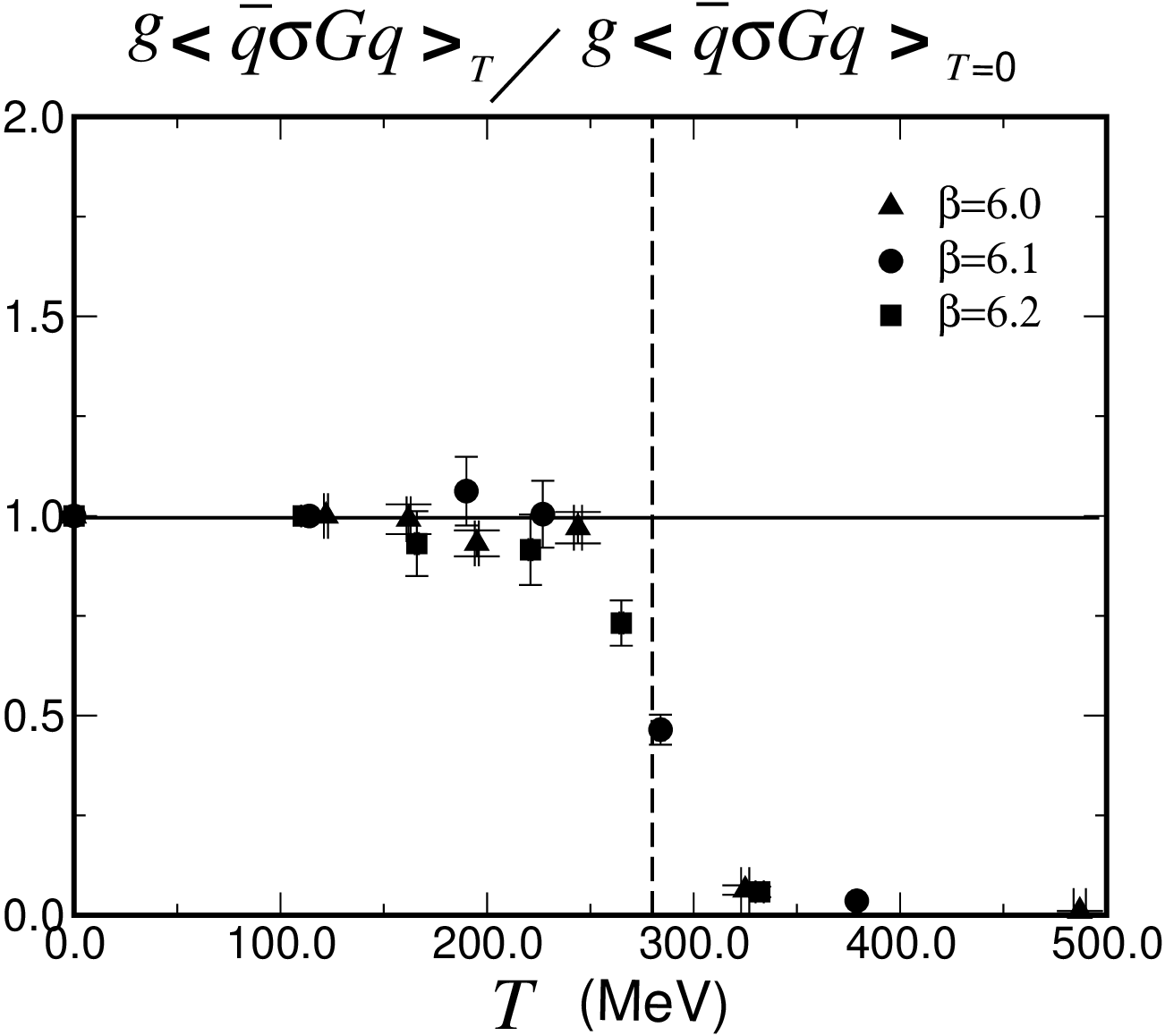}
\caption{
The quark-gluon mixed condensate $\qGq_T$ at finite temperature $T$ 
normalized by $\qGq_{T=0}$.
The vertical dashed line denotes the critical 
temperature $T_c \simeq 280 {\rm MeV}$
in quenched QCD.
}
\label{fig:qGq_finite_T}
\end{figure}

\begin{figure}[htb]
\centering
\includegraphics[scale=0.24]
{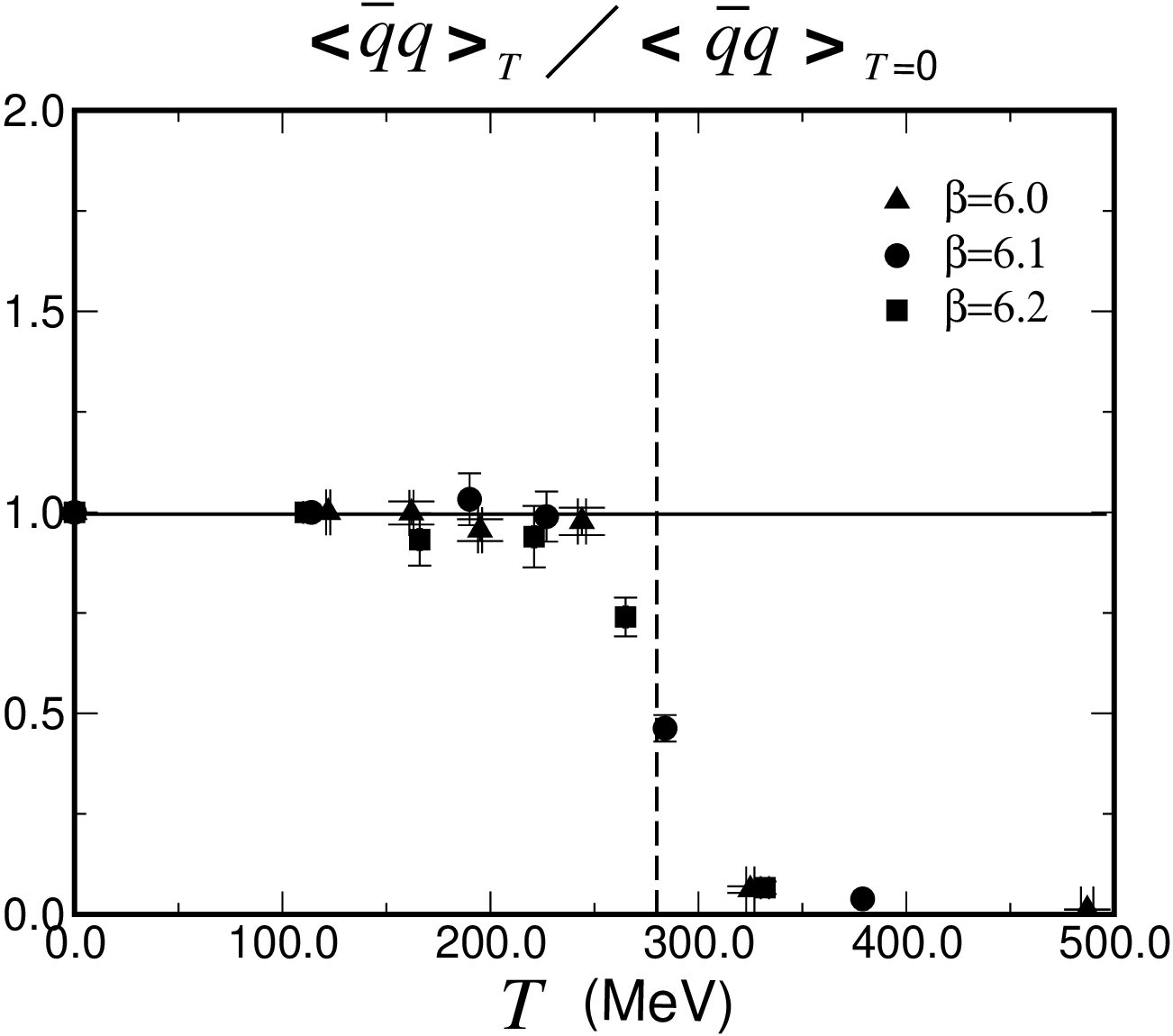}
\caption{
The quark condensate $\qq_T$ at finite temperature $T$ 
normalized by $\qq_{T=0}$.
The vertical dashed line denotes $T_c$.
}
\label{fig:qq_finite_T}
\end{figure}

%
In figure~\ref{fig:qGq_finite_T}, we plot the thermal effects on 
$\qGq$.
We find a drastic change of $\qGq$ around the critical temperature
$T_c \simeq 280 {\rm MeV}$. This is the first observation of 
chiral-symmetry restoration  through  $\qGq$.
We obtain $T_c/\sqrt{\sigma} = 0.64(4)$, which 
is consistent with the coincidence 
of the confinement/deconfinement phase transition 
and chiral-symmetry phase restoration.
We also find that 
the thermal effects on $\qGq$ are remarkably weak
below $T_c$,
namely, $T \simleq 0.9\, T_c$.
In figure~\ref{fig:qq_finite_T}, 
the thermal effects on $\qq$ are plotted,
and the same features as $\qGq$ are obtained.
%
%

\begin{figure}[htb]
\centering
\includegraphics[scale=0.24]
{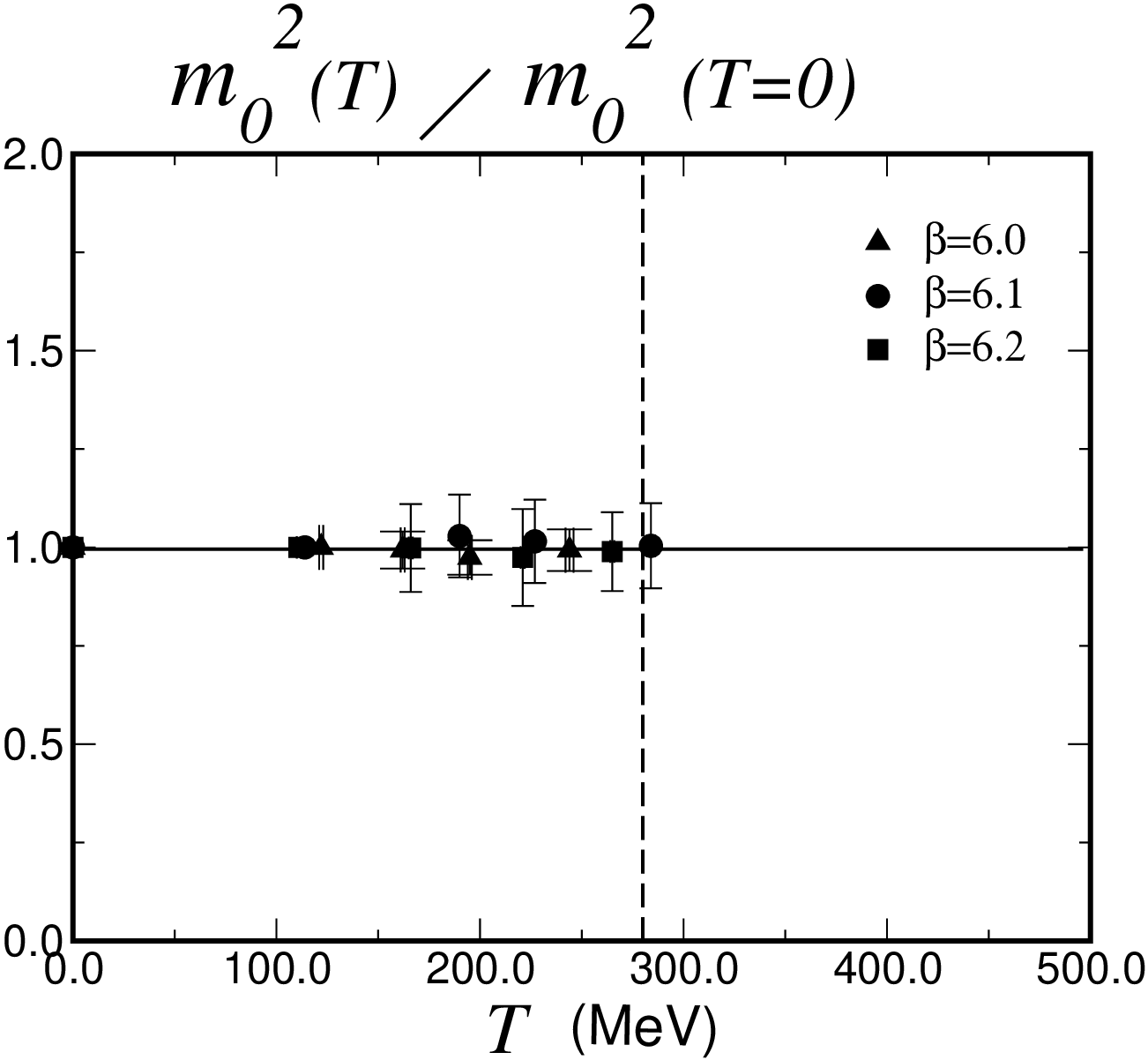}
\caption{The ratio $m_0^2(T) \equiv \qGq_T/\qq_T$ 
normalized by $m_0^2(T=0)$ 
plotted against $T$.
This result indicates 
the same chiral behavior 
between $\qGq_T$ and $\qq_T$.
}
\label{fig:M0_finite_T}
\end{figure}

We then quantitatively compare 
the thermal effects of $\qGq$ and $\qq$.  
For this purpose,
we plot 
the ratio $m_0^2(T) \equiv \qGq_T / \qq_T$ normalized by $m_0^2(T=0)$ 
for $T \simleq T_c$
in figure~\ref{fig:M0_finite_T}.
%
%
%
%
We observe that $m_0^2(T)$ is almost 
independent of the temperature, even in the very vicinity of $T_c$, which 
can be interpreted
that $\qGq_T$ and $\qq_T$ obey 
the same thermal behavior.
This result is rather nontrivial because, as was noted before, 
these two condensates characterize different aspects of the QCD vacuum.

As  a new  interesting feature  absent in  the  scalar-type condensate
$\qq$, $\qGq$  may reveal possible breaking of  the space-time duality
in Euclidean QCD at finite temperature.
For further analysis in this direction, 
we investigate the thermal effects on 
electric/magnetic components of $\qGq$ separately, i.e.,
$\qGqE$ and $\qGqB$.
Hereafter, no summation is taken for $i,j,k \in \{1,2,3\}$.
Of course, the equality $\qGqE = \qGqB$ holds at zero temperature from
O(4)  symmetry in  Euclidean QCD.   This equality,  however,  does not
necessarily hold for finite  temperature, where the space-time duality
is explicitly broken.
In fact, at finite temperature, $\qGqE_T$ and $\qGqB_T$ can be 
regarded as different chiral order parameters.

\begin{figure}[htb]
\centering
\includegraphics[scale=0.24]
{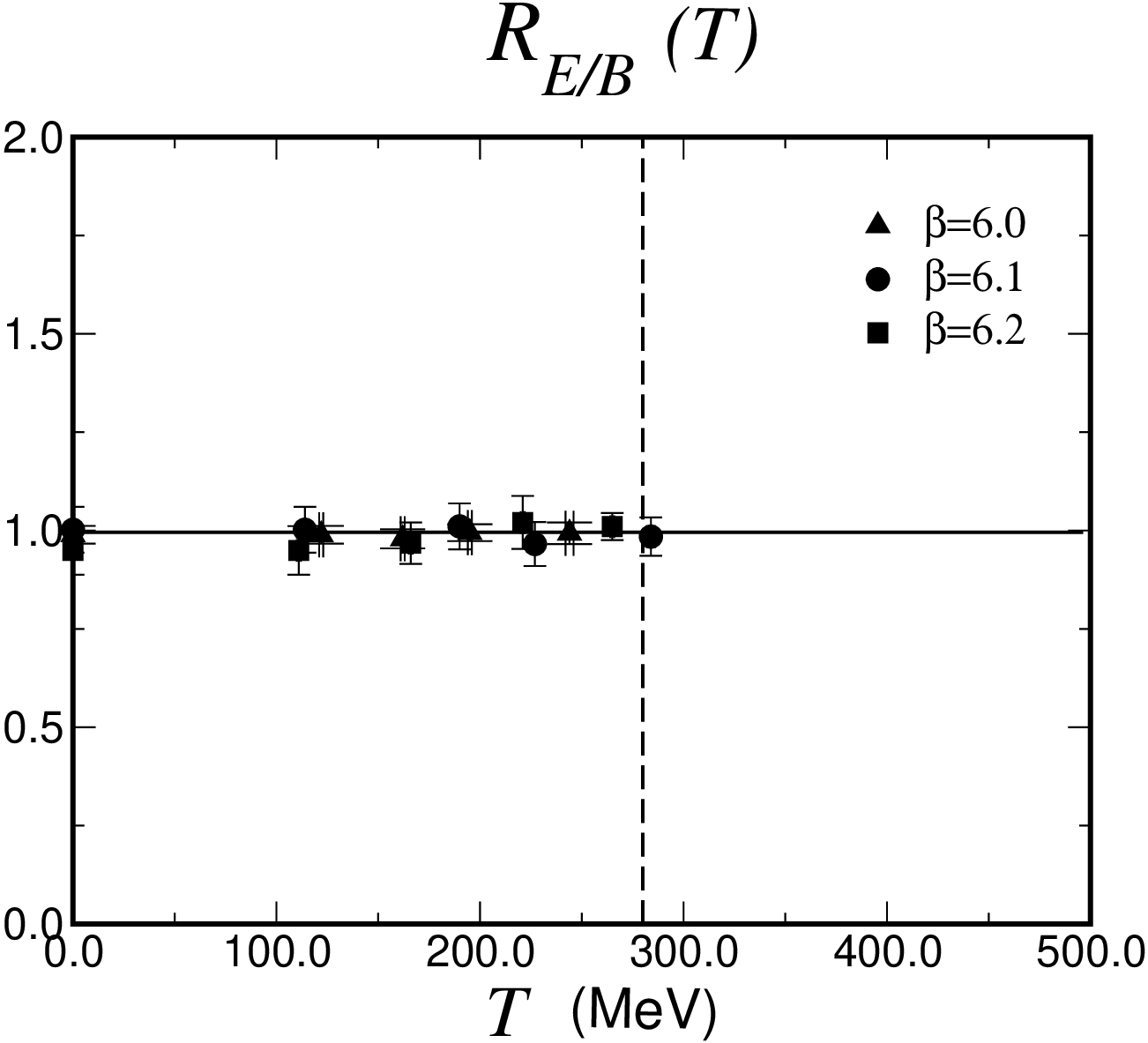}
\caption{
The ratio $R_{E/B}(T) \equiv \sum_i \qGqE_T / \sum_{j<k}\qGqB_T$ 
plotted against the temperature $T$.
This result indicates
$\qGqE$ and $\qGqB$ obey 
the same chiral behavior.
}
\label{fig:qGq_B/E_finite_T}
\end{figure}

We therefore plot the ratio $R_{E/B}(T)$ 
defined by
\begin{eqnarray}
R_{E/B}(T) \equiv \frac{\sum\limits_i \qGqE_T}{\sum\limits_{j<k}\qGqB_T}
\end{eqnarray}
at      finite      temperature      $T      \simleq      T_c$      in
figure~\ref{fig:qGq_B/E_finite_T}.
One sees that 
the equality $\qGqE_T = \qGqB_T$ holds for even at finite temperature.
This again indicates that
all the $(\mu,\nu)$ components of $\qGq_T$ obey 
the same chiral behavior near $T_c$. 
%
%

In this  way, we observe  that the chiral condensates,  i.e., $\langle
\bar{q}q\rangle_T$,   $g\langle  \bar{q}   \sigma_{\mu\nu}  G_{\mu\nu}
q\rangle_T$,    $g\langle   \bar{q}\sigma_{4i}G_{4i}q\rangle_T$   and
$g\langle  \bar{q}  \sigma_{jk}G_{jk}  q\rangle_T$,  obey  the  almost
identical behavior near $T_c$.
One of  the possible explanations of  this similarity is  given by the
so-called ``Swiss cheese picture''.   In this picture, phase structure
of the  QCD vacuum at the  critical temperature is  represented by the
mixture  of  co-existing two  phases,  i.e.,  chiral-broken phase  and
chiral-symmetric phase.   Since the deconfinement  phase transition is
of the first order in  the quenched approximation, the co-existence of
two phases is plausible.
In  this  scenario,  the   common  thermal  behaviors  of  the  chiral
condensates  can  be  explained  if  the thermal  effects  are  mainly
characterized by  the volume ratio of  the two phases  and the thermal
effects on each phase are subdominant.
Since  this consideration can  be applied  to any  chiral condensates,
this  scenario  predicts a  universal  behavior  of  all chiral  order
parameters  near  $T_c$,  which  provides  new aspects  on  the  chiral
structure of the QCD phase transition.

The common  thermal behaviors  of the chiral  condensates can  be also
explained      from     a     different      viewpoint.      Following
Refs.~\cite{Banks,Hands}, we can express the condensates as
\begin{eqnarray}
\begin{array}{ccl}
\qq &=&
\displaystyle\frac{1}{V}\int d\lambda' \frac{m \rho(\lambda')}{\lambda'^2 + m^2}, \\[3mm]
\qGq &=&
\displaystyle\frac{1}{V}\int d\lambda' \frac{m \rho(\lambda')}{\lambda'^2 + m^2}
\braket{\lambda'|\sigma_{\mu\nu}G_{\mu\nu}|\lambda'},
\end{array}
\end{eqnarray}
where $|\lambda\ket$ denotes the eigenvector of the Dirac operator
as $i\fslash{D} |\lambda\ket = \lambda |\lambda\ket$, and 
$\rho(\lambda)$ the spectral density on $\lambda$.
When one takes the chiral limit $m\rightarrow 0$ after the 
thermodynamic limit $V\rightarrow 0$, the zero modes ($\lambda=0$) 
of the Dirac operator are responsible for both the condensates.
If $\braket{\lambda|\sigma_{\mu\nu}G_{\mu\nu}|\lambda}$ 
does not have a singularity at $\lambda=0$, $m_0^2$ corresponds to
$\braket{\lambda|\sigma_{\mu\nu}G_{\mu\nu}|\lambda}\Big|_{\lambda=0}$.
The universal chiral behavior against the temperature
indicates that both the electric and the magnetic
components of 
$\braket{\lambda|\sigma_{\mu\nu}G_{\mu\nu}|\lambda}\Big|_{\lambda=0}$
for the chiral zero modes 
have remarkably weak dependence on the temperature 
for $T \simleq T_c$, even in the very vicinity of $T_c$.


\section{Summary}
\label{sec:summary}

We have studied thermal effects on
$\qGq$, in comparison with $\qq$,
using the SU(3)$_c$ lattice QCD with  
the KS-fermion at the quenched level.
For $T \simleq 0.9\, T_c$, both the condensates show very little
thermal effect, while a clear signal of chiral restoration is 
observed  near $T_c$ as a sharp decrease of both the condensates.
It has been a surprise that the ratio
$m_0^2(T) \equiv \qGq_T /\qq_T$, in contrast to 
the drastic change of each,
is almost independent of the temperature in the entire 
region of $T$ up to $T_c$.
It is also found that the Lorentz components, 
$\qGqE_T$ and $\qGqB_T$, show
an identical temperature dependence.
Such observations indicate that these chiral condensates show
a common thermal behavior, which
may reveal new aspects of the chiral restoration of the QCD vacuum.
For further studies,
a full QCD lattice calculation is in progress 
in order to analyze the dynamical quark effects on the condensates.
Through the full QCD calculation and its comparison
with quenched QCD, we expect to get deeper insight on the 
chiral structure of the QCD vacuum at finite temperature.


\acknowledgments
This work is supported in part by the Grant for Scientific Research 
( (B) No.15340072 and No.13011533) 
from the Ministry of Education, Culture, Science and Technology, Japan.
T.D. acknowledges the support by the JSPS
(Japan Society for the Promotion of Science) Research Fellowships 
for Young Scientists.
The Monte Carlo simulations have been performed on the 
NEC SX-5 supercomputer at Osaka University and 
IBM POWER4 Regatta system at Tokyo Institute of Technology.






\begin{thebibliography}{99}

%

\bibitem{Savvidy&Nielsen}     {
G.K.~Savvidy, Phys. Lett. {\bf B71}, 133 (1977);
N.K.~Nielsen and  P.~Olesen, Nucl. Phys. {\bf B144}, 376 (1978).}


\bibitem{Narison}         {S.~Narison, ``QCD Spectral Sum Rules", (World Scientific, 1989) p.1 and references therein.}

\bibitem{Ioffe}     {B.L.~Ioffe,
                            Nucl. Phys.  {\bf B188}, 317 (1981)
			    {\it ibid.} {\bf B191}, 591 (1981).}

\bibitem{Dosch}     
{H.G.~Dosch, M.~Jamin, and S.~Narison,
 Phys. Lett. {\bf B220}, 251 (1989);
W-Y.P.~Hwang and K.-C.~Yang,
 Phys. Rev. {\bf D49}, 460 (1994). }

\bibitem{Jido:parity} {D.~Jido, N.~Kodama and M.~Oka, 
			Phys. Rev. {\bf D54}, 4532 (1996);
			D. Jido and M. Oka, hep-ph/9611322.}

\bibitem{penta1}     {J.~Sugiyama, T.~Doi and M.~Oka,
                        Phys. Lett. {\bf B581}, 167 (2004).}

\bibitem{Nakano}     {T.~Nakano, et al., Phys. Rev. Lett. {\bf 91}, 012002 (2003).}

\bibitem{Dosch2}     {H.G.~Dosch, and S.~Narison,
                        Phys. Lett. {\bf B417}, 173 (1998)
			and references therein.}

\bibitem{Latorre}    {J.I.~Latorre, P.~Pascual, and S.~Narison,
			Z. Phys. {\bf C34}, 347 (1987).}

\bibitem{Bel}       {V.M.~Belyaev and B.L.~Ioffe,
			Sov. Phys. JETP {\bf 56}, 493 (1982).}

\bibitem{K&S}       {M.~Kremer and G.~Schierholz,
			Phys. Lett. {\bf B194}, 283 (1987).}

\bibitem{DOIS:qGq}  {T.~Doi, N.~Ishii, M.~Oka and H.~Suganuma,
                        Phys. Rev. {\bf D67}, 054504 (2003).}

\bibitem{twc:qGq}   {T.W.~Chiu and T.H.~Hsieh,
			Nucl. Phys. {\bf B673}, 217 (2003).}

\bibitem{DOIS:T}  {T.~Doi, N.~Ishii, M.~Oka and H.~Suganuma,
                        Nucl. Phys. {\bf A721}, 934 (2003);
                        Prog. Theor. Phys. Suppl. {\bf 151}, 161 (2003);
%
%
			Nucl. Phys. {\bf B129-130} (Proc. Suppl.), 566 (2004);
%
		T.~Doi, H.~Suganuma, M.~Oka and N.~Ishii,
			Proc. of 
			``Color Confinement and Hadrons in Quantum Chromodynamics
			  (Confinement 2003)'',
			(World Scientific, 2004) p.398, hep-lat/0311015.}



\bibitem{Gockeler:scale}    {
M.~Gockeler, R.~Horsley, H.~Perlt, 
P.~Rakow, G.~Schierholz, A.~Schiller
and P.~Stephenson,
			Phys. Rev. {\bf D57}, 5562 (1998).}

\bibitem{Boyd}	      {
G.~Boyd, J.~Engels, F.~Karsch, E.~Laermann, 
C.~Legeland, M.~Lutgemeier and B.~Petersson,
			Nucl. Phys. {\bf B469}, 419 (1996); 
N.~Ishii, H.~Suganuma and H.~Matsufuru, Phys. Rev. {\bf D66}, 094506 (2002).}

\bibitem{Narison2}  { S.~Narison and R.~Tarrach,
			Phys. Lett. {\bf B125}, 217 (1983).}


\bibitem{Banks}       {T.~Banks and A.~Casher,
			Nucl. Phys. {\bf B169}, 103 (1980).}

\bibitem{Hands}		{S.~Hands, J.B.~Kogut and A.~Kocic,
			Nucl. Phys. {\bf B357}, 467 (1991).}




\end{thebibliography}
\end{document}